\documentclass[12pt,a4paper]{article}
\usepackage{mathtools}
\usepackage{mathrsfs}
\usepackage{amsmath}
\usepackage{tikz-feynman}
\usepackage[utf8x]{inputenc}
\usepackage{amssymb}
\usepackage{slashed}
\usepackage{bbm}
\usepackage{bm}
\usepackage{color}
\usepackage[a4paper,top=3cm,bottom=3cm,left=2cm,right=3cm,bindingoffset=5mm]{geometry}
\usepackage{booktabs} % To thicken table lines
\usepackage{tikz}
\usepackage{cancel}
\usepackage[bottom]{footmisc}
\usepackage{cite}
\usepackage{float}
\usepackage{todonotes}
\usepackage{jheppub}
\usepackage[flushleft]{threeparttable}
\usepackage{makecell,booktabs}

\usepackage{hyperref}

\hypersetup{
colorlinks=true,         % false: boxed links; true: colored links, false is default
linkcolor=blue,          % color of internal links, red is default
citecolor=red,        % color of links to bibliography
urlcolor=red            % color of external links, cyan is default
}

\renewcommand{\[}{\begin{equation}\begin{aligned}}
\renewcommand{\]}{\end{aligned}\end{equation}}

\newcommand{\wb}{{\bar{w}}}

\newcommand{\ta}{\text{a}}
\newcommand{\tb}{\text{b}}
\newcommand{\tc}{\text{c}}

\title{From Moyal deformations to chiral higher-spin theories and to celestial algebras}
\author{Ricardo Monteiro}

\affiliation{Centre for Theoretical Physics, Department of Physics and Astronomy, \\
Queen Mary University of London, E1 4NS, United Kingdom}

\abstract{
We study the connection of Moyal deformations of self-dual gravity and self-dual Yang-Mills theory to chiral higher-spin theories, and also to deformations of operator algebras in celestial holography. The relation to Moyal deformations illuminates various aspects of the structure of chiral higher-spin theories. For instance, the appearance of the self-dual kinematic algebra in all the theories considered here leads via the double copy to vanishing tree-level scattering amplitudes. Regarding celestial holography, the Moyal deformation of self-dual gravity was recently shown to lead to the loop algebra of $W_{\wedge}$, and we obtain here the analogous deformation of a Kac-Moody algebra corresponding to Moyal-deformed self-dual Yang-Mills theory. We also introduce the celestial algebras for various chiral higher-spin theories. 
}

\begin{document}

\begin{flushright}
QMUL-PH-22-39
\end{flushright}

\maketitle

%%%%%%%%%%%%%%%%%%%%%%%%%%%%%%%%%%%%%%%%%%
%%%%%%%%%%%%%%%%%%%%%%%%%%%%%%%%%%%%%%%%%%

\section{Introduction}

This paper aims to connect three topics that have ignited interest at different times among theoretical physicists. The topics are: Moyal deformations of field theories, chiral higher-spin theories of massless particles, and chiral algebras that appear in celestial holography.

The Moyal bracket was introduced long ago to provide a geometric understanding of phase-space non-commutativity in quantum mechanics \cite{Moyal:1949sk}. It has since been applied also to describe non-commutativity in a position-space plane. It is characterised by a deformation parameter, which we will call $\alpha$, and it turns out to be the most general deformation of a Poisson bracket involving two coordinates that is still a Lie bracket \cite{Fletcher:1990ib}. Because of this property, it is often the case that when a Poisson bracket arises in an integrable system, the Moyal deformation of the bracket preserves the integrability. That was the original motivation to introduce a Moyal deformation of self-dual gravity, whose equation of motion is integrable \cite{STRACHAN199263}. Moyal deformations became the subject of intense interest in higher-energy theory when they appeared in M/string theory; see e.g.~\cite{Connes:1997cr,Nekrasov:1998ss,Schomerus:1999ug,Seiberg:1999vs} as well as the reviews \cite{Douglas:2001ba,Szabo:2001kg}. For instance, it was described in \cite{Seiberg:1999vs} how Yang-Mills theory in a space with a non-commutative plane arises from the low-energy limit of open strings on a background with a B-field. The Moyal deformation of self-dual Yang-Mills theory was considered in \cite{Nekrasov:1998ss}. Our formulation of the Moyal-deformed self-dual theories, closer to the works \cite{STRACHAN199263,Takasaki:2000vs,Garcia-Compean:2003nix}, focuses on the case where the plane of non-commutativity is null and can be chosen to align with a particular gauge choice. This case allows for very simple Moyal-deformed equations of motion of the self-dual theories in light-cone gauge.

The Moyal-deformed theories are not Lorentz invariant, due to the plane of non-commutativity. As we will see, however, if we interpret their fields as generating functions of higher-spin fields, these non-Lorentz-invariant theories generate Lorentz-invariant theories of massless higher-spin particles. The latter coincide with the `chiral higher-spin theories' discussed in \cite{Ponomarev:2016lrm,Ponomarev:2017nrr,Skvortsov:2018jea}, based on the chiral three-point vertices introduced in \cite{Metsaev:1991mt,Metsaev:1991nb}; see \cite{Ponomarev:2022vjb,Bekaert:2022poo} for related reviews. These theories are higher-spin extensions of self-dual gravity/Yang-Mills. Moyal brackets are known to arise in higher-spin theories, and for the particular chiral theories that we study here, the Moyal bracket was identified from the vertices in \cite{Ponomarev:2017nrr}; see also \cite{Skvortsov:2022syz,Tran:2022tft}. We develop further these insights by explicitly connecting the chiral higher-spin theories to Moyal-deformed self-dual gravity/Yang-Mills, and exploring some of the implications.

It is natural to expect that 4D-chiral theories lead to 2D-chiral celestial algebras, by which we mean the operator-product algebras that arise from the consideration of asymptotic scattering states in celestial holography; see \cite{Raclariu:2021zjz,Pasterski:2021rjz,Pasterski:2021raf,McLoughlin:2022ljp} for recent reviews. Celestial holography builds on the connection between asymptotic symmetries in asymptotically flat spacetimes and `soft theorems' obeyed by scattering amplitudes \cite{Strominger:2013jfa,He:2014laa,Strominger:2017zoo}. The soft tower associated to asymptotic symmetries \cite{Donnay:2018neh,Adamo:2019ipt,Puhm:2019zbl,Guevara:2019ypd} admits a very simple formulation when restricted to the self-dual sectors of gravity/Yang-Mills \cite{Guevara:2021abz}.  For self-dual gravity, one obtains (the loop algebra of the wedge subalgebra of) $w_{1+\infty}$ \cite{Strominger:2021lvk}. This algebra can be seen to arise in various ways; see \cite{Adamo:2021lrv} for a twistorial approach as well as a historical discussion of related results, or e.g.~refs.~\cite{Himwich:2021dau,Jiang:2021csc,Mago:2021wje,Freidel:2021ytz,Costello:2022wso,Costello:2022upu,Ren:2022sws,Compere:2022lzx,Monteiro:2022lwm}. Deformations of the chiral algebras of self-dual gravity/Yang-Mills have also been studied, with a focus on what class of deformations preserves the associativity of the algebras \cite{Himwich:2021dau,Mago:2021wje,Ren:2022sws,Melton:2022fsf}. It was found in \cite{Monteiro:2022lwm,Bu:2022iak} that the Moyal deformation of self-dual gravity leads to a known deformation of $w_{1+\infty}$ into $W_{1+\infty}$. The present paper was partly motivated by \cite{Ren:2022sws}, where it was observed in examples that the vertices of chiral higher-spin theories lead to an associative celestial algebra. See refs.~\cite{Ponomarev:2022ryp,Ponomarev:2022qkx} for other work on the celestial holography of chiral higher-spin theories.

There is a thread running through all these topics that illuminates their relationship, and that is the double-copy structure of the theories in question. The double copy is a factorisation of the interactions that most famously connects the perturbative structure of Yang-Mills theory and gravity; see recent reviews in \cite{Bern:2019prr,Bern:2022wqg,Kosower:2022yvp,Adamo:2022dcm} and the seminal papers \cite{Kawai:1985xq,Bern:2008qj}. For works relating celestial holography and the double copy, see \cite{Pasterski:2020pdk,Casali:2020vuy,Casali:2020uvr,Campiglia:2021srh,Godazgar:2021iae,Adamo:2021dfg,Gonzo:2022tjm,Huang:2019cja,Alawadhi:2019urr,Banerjee:2019saj,Monteiro:2022lwm,Guevara:2022qnm,Nagy:2022xxs}. In particular, refs.~\cite{Monteiro:2022lwm,Guevara:2022qnm} identified the property of associativity in chiral algebras, including the appearance of $w_{1+\infty}$ in self-dual gravity, with the algebraic structure of the colour-kinematics duality, which is very simple in the self-dual sector \cite{Monteiro:2011pc}. This extends to the Moyal deformations and also applies, as we will discuss here building on \cite{Ponomarev:2017nrr}, to the chiral higher-spin theories.

The structure of this paper is as follows. In section~\ref{sec:chsfromMoyal}, we describe how Moyal-deformed self-dual gravity/Yang-Mills generate chiral theories of higher-spin fields. In section~\ref{sec:doublecopy}, we discuss the related double-copy structures of Moyal deformations and of chiral higher-spin theories, and the vanishing of the tree amplitudes that is associated to classical integrability. The celestial chiral algebras arising from the various theories considered are presented in section~\ref{sec:celestial}. Section~\ref{sec:loop} provides a brief overview of loop amplitudes in these theories. The paper concludes with a discussion in section~\ref{sec:conclusion}.

%%%%%%%%%%%%%%%%%%%%%%%%%%%%%%%%%%%%%%%%%%
%%%%%%%%%%%%%%%%%%%%%%%%%%%%%%%%%%%%%%%%%%

\section{Chiral higher-spin theories from Moyal deformations}
\label{sec:chsfromMoyal}

In this section, we discuss the use of Moyal deformations of self-dual gravity and self-dual Yang-Mills theory as generating theories for the chiral higher-spin theories developed in \cite{Metsaev:1991mt,Metsaev:1991nb,Ponomarev:2016lrm,Skvortsov:2018jea}, and other related chiral higher-spin theories. Our light-cone-gauge approach is closely related to covariant approaches followed in \cite{Krasnov:2021nsq,Tran:2021ukl,Skvortsov:2022syz,Sharapov:2022faa,Adamo:2022lah,Sharapov:2022wpz,Sharapov:2022nps,Hahnel:2016ihf,Adamo:2016ple} and especially \cite{Tran:2022tft}, and builds on observations in \cite{Ponomarev:2017nrr} for the symmetry algebra of the interactions. We follow the notation of \cite{Monteiro:2022lwm}, which is useful for connecting the discussion to the literature on the double copy later in the paper.

%%%%%%%%%%%%%%%%%%%%%%%%%%%%%%%%%%%%%%%%%%

\subsection{chs($\alpha$) from Moyal-SDG}

The Moyal deformation of self-dual gravity (Moyal-SDG) \cite{STRACHAN199263} is defined in terms of the light-cone gauge formulation of SDG \cite{Plebanski:1975wn,Siegel:1992wd}. We employ light-cone coordinates $(u,v,w,\wb)$, with the wave operator $\,\square:=2(-\partial_u\partial_v+\partial_w\partial_\wb)\,$. The equation of motion of Moyal-SDG is\footnote{For notational simplicity, we suppress the gravitational coupling constant $\kappa$, which multiplies the interaction term. This is achieved by a rescaling of $\phi$. All the gravity-like theories in this paper inherit this coupling constant. The analogous statement is true for the Yang-Mills-like theories to be seen later.}
\[
\label{eq:MSDG}
\square \phi + \{\partial_u \phi,\partial_w \phi\}^M =0 \,.
\] 
The difference with respect to SDG is that the latter's Poisson bracket
\[
\label{eq:PoissonP}
\{f,g\}:=f \buildrel{\leftrightarrow} \over {P} g \,, \qquad 
\buildrel{\leftrightarrow} \over {P} \; = \;
\buildrel{\leftarrow} \over{\partial}_u
\buildrel{\rightarrow} \over{\partial}_w-
\buildrel{\leftarrow} \over{\partial}_w
\buildrel{\rightarrow} \over{\partial}_u\,,
\]
is substituted by its deformation into the Moyal bracket
\[
\label{eq:Moyal}
\{f,g\}^M:=\frac{1}{2\alpha}(f\star g-g\star f) =\frac{1}{\alpha}\, f \sinh(\alpha \! \buildrel{\leftrightarrow} \over {P})\, g \,,
\]
defined from the Moyal product \,$f\star g:= f\exp(\alpha \! \buildrel{\leftrightarrow} \over {P})\, g$\,, which is associative but not commutative. The parameter $\alpha$ characterises the deformation, such that \,$\{f,g\}^M \to \{f,g\}$\, as \,$\alpha\to0$\,.
The Moyal bracket is the most general Lie bracket of functions of two variables \cite{Fletcher:1990ib}, here $(u,w)$.\footnote{Notice that the chosen non-commutative plane is complex in Lorentzian signature. In fact, so are all the chiral-type theories we consider in this paper. A more natural setting is split signature, where the coordinates $w$ and $\bar w$ are real and independent, and so are the two chiralities of the fields, e.g.~$\phi_h$ and $\phi_{-h}$ to be seen below, where $h$ is the helicity.}

It was suggested in \cite{Monteiro:2022lwm} that $\phi$ may be interpreted as a composite field of chiral higher-spin fields. This is closely related to discussions in \cite{Ponomarev:2017nrr} and also to a large body of work concerning the symmetries of higher-spin theories. Here, we aim to make that suggestion more concrete. Suppose that we write
\[
\label{eq:phialpha}
\phi = \zeta^2  \sum_{h\in {\mathbb Z}} \zeta^{-h}\, \phi_h\,,
\]
where $\phi_h$ is a helicity-$h$ field and $ \zeta$ is a parameter that will allow us to project into specific helicity spaces. Now, consider the Laurent expansion of \eqref{eq:MSDG} in powers of $\zeta$ after substituting $\alpha \mapsto \alpha \,\zeta\, $,
\begin{align}
\label{eq:MSDGalpha}
0&=\square \phi + \phi\,  \frac{\buildrel{\leftrightarrow} \over {P}  \sinh(\alpha\,\zeta \! \buildrel{\leftrightarrow} \over {P})}{\alpha\,\zeta} \, \phi =\square \phi + \phi\, \Bigg(\, \sum_{\sigma\geq1}  \frac{(\alpha\,\zeta\buildrel{\leftrightarrow} \over {P})^{2\sigma}}{\alpha^2\,\zeta^2\,(2\sigma-1)!}\, \Bigg)\, \phi \nonumber\\
&= \zeta^2\, \sum_{h} \zeta^{-h}\left(\square\phi_h + \!\!\sum_{\substack{h_1,h_2 \\ \text{even}\;h_1+h_2-h>0}} \!\!\phi_{h_1}\; \frac{(\alpha\buildrel{\leftrightarrow} \over {P})^{h_1+h_2-h}}{\alpha^2\,(h_1+h_2-h-1)!}\; \phi_{h_2} \right) \,.
\end{align}
Imposing that the equations of motion hold independently for each coefficient in the $\zeta$-expansion,
we obtain precisely the equations of motion of the chiral higher-spin theory of \cite{Ponomarev:2016lrm,Metsaev:1991mt,Metsaev:1991nb}, which we will name chs($\alpha$).\footnote{The helicity fields in these references are translated into our conventions as $(\partial_u)^h\,\phi_{h}$ in order to simplify the expressions. This causes the interactions to appear to have higher order in derivatives, e.g.~for self-dual gravity in the first line of \eqref{eq:SDGalpha}.\label{foot:partialu}} In fact, we can also obtain the action for this theory, by introducing a projector ${\mathcal P}_{(\zeta)}$ into the helicity-0 part that acts linearly as
\[
\label{eq:P}
{\mathcal P}_{(\zeta)}\,\zeta^n = \delta_{n,0} \,\qquad \forall n\in {\mathbb Z}\,,
\]
and writing\footnote{It would appear to be more convenient here to define ${\mathcal P}_{(\zeta)}\,\zeta^n = \delta_{n,4}$, but other choices are more convenient for later examples. So we chose instead to make the definition \eqref{eq:P} and multiply by $\zeta^{-4}$ in the first line. Any such definition can be enforced via a contour integral.}
\begin{align}
\label{eq:actionchs}
S_{\text{chs}(\alpha)}(\phi_h)
&={\mathcal P}_{(\zeta)} \int d^4 x \Bigg( \frac1{2}\, \phi\,\square\,\phi + \frac1{3}\,  \frac{\left(\phi\buildrel{\leftrightarrow} \over {P}  \sinh(\alpha\,\zeta \! \buildrel{\leftrightarrow} \over {P})\,\phi\right)}{\alpha\,\zeta} \; \phi\Bigg) \zeta^{-4}
\nonumber \\
&=\int d^4 x \Bigg( \frac1{2}\sum_h \phi_{-h}\,\square\,\phi_h \,+ \frac1{3}\!\!\sum_{\substack{h_1,h_2,h_3 \\ \text{even}\;h_1+h_2+h_3>0}} \!\! \frac{\big(\phi_{h_1}\big(\alpha\buildrel{\leftrightarrow} \over {P}\big)^{h_1+h_2+h_3}\phi_{h_2}\big)}{\alpha^2\,(h_1+h_2+h_3-1)!}\;\phi_{h_3}\Bigg)  \,.
\end{align}
An important feature is that while Moyal-SDG is not a Lorentz-invariant theory (due to the plane of non-commutativity being special), the chiral higher-spin theory it generates is in fact Lorentz invariant. We discuss this point further in section~\ref{subsec:3ptamp}.

Simpler versions of a chiral higher-spin theory have been considered in the literature. It is possible to restrict the theory above to admit only even spins. Independently, it is possible to simplify the interactions. Suppose that we apply \eqref{eq:phialpha} to SDG, rather than to its Moyal deformation:
\begin{align}
\label{eq:SDGalpha}
0&=\square \phi + \phi\,  \buildrel{\leftrightarrow} \over {P} {}^2 \, \phi \nonumber\\
&= \zeta^2 \,\sum_{h} \zeta^{-h}\Bigg(\square\phi_h + \!\!\sum_{\substack{h_1,h_2 \\ h_1+h_2-h=2}} \!\!\phi_{h_1}\, \buildrel{\leftrightarrow} \over {P} {}^2\, \phi_{h_2} \Bigg) \,.
\end{align}
This gives a two-derivative chiral higher-spin theory introduced in \cite{Ponomarev:2017nrr} as a `contraction' of the more involved theory of \cite{Ponomarev:2016lrm} we saw above. The action can be obtained analogously to \eqref{eq:actionchs}, keeping only the interaction terms obeying $h_1+h_2+h_3=2$\,. In fact, naming the theory with action \eqref{eq:actionchs} chs($\alpha$), this simpler theory is just chs($0$). So even undeformed SDG generates via \eqref{eq:phialpha} a theory of higher spins.

We can also consider an even simpler chiral higher-spin theory, which was called `higher-spin SDG' in \cite{Krasnov:2021nsq}. While the theories above are all higher-spin generalisations of SDG in some sense, this theory has the closest structure, because it possesses only $(+^{s_1}+^{s_2}-^{s_3})$ vertices, for spins $s_i\geq2$; we recall that SDG has a single vertex $(+^{2}+^{2}-^{2})$. In this case, we work with the generating fields
\[
\label{eq:phibarphialpha}
\phi = \zeta^2  \sum_{s\geq2} \zeta^{-s}\, \phi_s\,,
\qquad\quad 
\bar\phi = \zeta^{-2}  \sum_{s\geq2} \zeta^{s}\, \phi_{-s}\,,
\]
to obtain the action
\begin{align}
\label{eq:actionhsSDG}
S_\text{hsSDG}(\phi_{\pm s}|_{s\geq2})
&={\mathcal P}_{(\zeta)} \int d^4 x \;\; \bar\phi\,\left( \square \phi + \phi\,  \buildrel{\leftrightarrow} \over {P} {}^2 \, \phi \right)
\nonumber \\
&=\int d^4 x \sum_{s\geq2} \phi_{-s} \Bigg(\square\phi_s + \!\!\sum_{\substack{s_1,s_2\geq2 \\ s_1+s_2-s=2}} \!\!\phi_{s_1}\, \buildrel{\leftrightarrow} \over {P} {}^2\, \phi_{s_2} \Bigg)   \,.
\end{align}

%%%%%%%%%%%%%%%%%%%%%%%%%%%%%%%%%%%%%%%%%%

\subsection{gl-chs($\alpha$) from Moyal-SDYM}

Here we consider the `gluonic' chiral higher-spin theory, with U(N)-valued fields, that is associated to Moyal-deformed self-dual Yang-Mills theory (Moyal-SDYM). The equation of motion of Moyal-SDYM is \cite{Takasaki:2000vs,Nekrasov:1998ss}
\[
\label{eq:MSDYM}
\square \Psi + [\partial_u \Psi,\partial_w \Psi]^M=0 \,.
\]
For U(N)-valued $A$ and $B$, we define
\[
[A,B]^M := A \star B - B \star A = A \,\exp(\alpha \! \buildrel{\leftrightarrow} \over {P})\, B - B\, \exp(\alpha \! \buildrel{\leftrightarrow} \over {P})\, A \; \stackrel{\alpha\to0}{\longrightarrow}\;  [A,B]\,.
\]
The equation \eqref{eq:MSDYM} can be written as
\[
\label{eq:MSDYMa}
\square \Psi^\ta + f^{\ta\tb\tc}\,\Psi^\tb \buildrel{\leftrightarrow} \over {P}\cosh(\alpha \! \buildrel{\leftrightarrow} \over {P})\, \Psi^\tc + d^{\ta\tb\tc}\,\Psi^\tb \buildrel{\leftrightarrow} \over {P}\sinh(\alpha \! \buildrel{\leftrightarrow} \over {P})\, \Psi^\tc =0\,,
\]
where we denote
\[
T^\ta T^\tb - T^\tb T^\ta = f^{\ta\tb\tc} \,T^\tc\,, \qquad T^\ta T^\tb + T^\tb T^\ta = d^{\ta\tb\tc} \,T^\tc\,.
\]

Analogously to \eqref{eq:phialpha}, we consider the generating field
\[
\label{eq:Psialpha}
\Psi = \zeta \sum_{h=-\infty}^\infty \zeta^{-h}\, \Psi_h\,,
\]
where $\Psi_h$ is a helicity-$h$ field. The Laurent expansion of \eqref{eq:MSDYM} in powers of $\zeta$ after substituting $\alpha \mapsto \alpha \,\zeta\, $ is
\begin{align}
\label{eq:MSDYMalpha}
0&=\square \Psi^\ta + f^{\ta\tb\tc}\,\Psi^\tb \buildrel{\leftrightarrow} \over {P}\cosh(\alpha\, \zeta\! \buildrel{\leftrightarrow} \over {P})\, \Psi^\tc + d^{\ta\tb\tc}\,\Psi^\tb \buildrel{\leftrightarrow} \over {P}\sinh(\alpha\,\zeta \! \buildrel{\leftrightarrow} \over {P})\, \Psi^\tc \nonumber\\
&= \zeta \sum_{h} \zeta^{-h}\Bigg(\square\Psi_h^\ta + \!\sum_{\substack{h_1,h_2 \\ h_1+h_2-h>0}} \!\!t^{\ta\tb\tc}_{h_1+h_2-h}\, \frac{ \Psi_{h_1}^\tb\,\big(\alpha\!\buildrel{\leftrightarrow} \over {P}\big)^{h_1+h_2-h}\, \Psi_{h_2}^\tc}{\alpha\,(h_1+h_2-h-1)!} \Bigg) \,,
\end{align}
where we define, for convenience,
\[
\label{eq:deft}
t^{\ta\tb\tc}_{\Sigma h} = f^{\ta\tb\tc} \;\;  \text{for odd}\; \Sigma h\,, \qquad t^{\ta\tb\tc}_{\Sigma h} = d^{\ta\tb\tc} \;\;  \text{for even}\; \Sigma h\,.
\]
We obtain the equations of motion of the gluonic chiral higher-spin theory considered in \cite{Skvortsov:2018jea,Skvortsov:2020wtf}. The action is obtained from
\begin{align}
\label{eq:actionglchs}
&S_{\text{gl-chs}(\alpha)}(\Psi_h)
={\mathcal P}_{(\zeta)} \int d^4 x\;\;\text{tr} \Bigg( \frac1{2}\, \Psi\,\square\,\Psi + \frac1{3}\,  \Psi\,[\partial_u \Psi,\partial_w \Psi]^M
\Bigg) \zeta^{-2}
\nonumber \\
&\quad =\int d^4 x\; \Bigg( \frac1{2}\sum_h \Psi^\ta_{-h}\,\square\,\Psi^\ta_h \,+ \frac1{3}\!\!\sum_{\substack{h_1,h_2,h_3 \\ h_1+h_2+h_3>0}} \!\! t^{\ta\tb\tc}_{h_1+h_2+h_3}\,\frac{ \big( \Psi_{h_1}^\ta\,\big(\alpha\!\buildrel{\leftrightarrow} \over {P}\big)^{h_1+h_2+h_3}\, \Psi_{h_2}^\tb\big)}{\alpha\,(h_1+h_2+h_3-1)!} \; \Psi_{h_3}^\tc\Bigg)  \,.
\end{align}

If we apply \eqref{eq:Psialpha} to SDYM, rather than to its Moyal deformation, we obtain
\begin{align}
\label{eq:SDYMalpha}
0&=\square \Psi^\ta + f^{\ta\tb\tc}\,\Psi^\tb \buildrel{\leftrightarrow} \over {P}\, \Psi^\tc  \nonumber\\
&= \zeta \sum_{h} \zeta^{-h}\Bigg(\square\Psi_h^\ta + f^{\ta\tb\tc} \!\!\sum_{\substack{h_1,h_2 \\ h_1+h_2-h>0}} \!\!\Psi_{h_1}^\tb\, \buildrel{\leftrightarrow} \over {P}\, \Psi_{h_2}^\tc \Bigg) \,,
\end{align}
If we call the theory \eqref{eq:actionglchs} gl-chs($\alpha$), then the theory with the equations of motion given in \eqref{eq:SDYMalpha} is simply gl-chs(0). It was first considered in \cite{Ponomarev:2017nrr}, and it can be truncated to admit only odd spins, unlike gl-chs($\alpha$).

Finally, again analogously to the gravity case, there is a theory even more similar to SDYM, which was called `higher-spin SDYM' in \cite{Krasnov:2021nsq}. The generating fields are
\[
\Psi = \zeta  \sum_{s\geq1} \zeta^{-s}\, \Psi_s\,,
\qquad\quad 
\bar\Psi = \zeta^{-1}  \sum_{s\geq1} \zeta^{s}\, \Psi_{-s}\,,
\]
and the action of this theory is
\begin{align}
\label{eq:actionhsSDYM}
S_\text{hsSDYM}(\Psi_{\pm s}|_{s\geq1})
&={\mathcal P}_{(\zeta)} \int d^4 x \;\; \text{tr} \;\; \bar\Psi \left( \square \Psi + [\partial_u \Psi,\partial_w \Psi] \right)
\nonumber \\
&=\int d^4 x \sum_{s\geq1} \Psi^\ta_{-s} \Bigg(\square\Psi_s^\ta + f^{\ta\tb\tc}\!\!\sum_{\substack{s_1,s_2\geq1 \\ s_1+s_2-s=1}} \!\!  \Psi^\tb_{s_1}\, \buildrel{\leftrightarrow} \over {P}\, \Psi^\tc_{s_2} \Bigg)   \,.
\end{align}

\subsection{Useful table of chiral higher-spin theories}

For the reader's convenience, we list the chiral higher-spin theories discussed above in table~\ref{tablechs}.

\begin{table}
  \caption{Table of chiral higher-spin theories. The generation technique is described in section~\ref{sec:chsfromMoyal}. The vertices are discussed in section~\ref{sec:doublecopy}. The actions appear in section~\ref{sec:chsfromMoyal}, but their relevance is discussed in section~\ref{sec:loop}. Truncations to even spins are allowed in chs($\alpha$), chs($0$) and hsSDG. Truncations to odd spins are allowed in gl-chs($0$) and hsSDYM. \\ \label{tablechs}}
  \vspace{.4cm}
  \centering
  \begin{threeparttable}
    \begin{tabular}{c@{\qquad}c@{\qquad}c}
    
      Chiral higher-spin theory & Vertices & Generating theory \\
      
       \midrule\midrule 
       
        \makecell{chs($\alpha$) \vspace{.2cm} \\   all spins} & 
        \makecell{$\displaystyle  \frac{\big(\alpha X(k_1,k_2)\big)^{h_1+h_2+h_3}}{\alpha^2(h_1+h_2+h_3-1)!}$  \vspace{.2cm} \\ even $h_1+h_2+h_3>0$} &  \makecell{Moyal-SDG \vspace{.2cm} \\   action($\phi$) } \\
     \cmidrule(l r){1-3}
     \makecell{chs($0$) \vspace{.2cm} \\    all spins} &  \makecell{$X(k_1,k_2)^2$  \vspace{.2cm} \\ $h_1+h_2+h_3=2$} &  \makecell{SDG \vspace{.2cm} \\  action($\phi$) } \\
    \cmidrule(l r){1-3}
      \makecell{hsSDG \vspace{.2cm} \\    $s\geq2$} & \makecell{$X(k_1,k_2)^2$  \vspace{.2cm} \\ $s_1+s_2-s_3=2$} &   \makecell{SDG \vspace{.2cm} \\  action($\phi,\bar\phi$) } \\ 
      
      \midrule\midrule
      
              \makecell{gl-chs($\alpha$) \vspace{.2cm} \\   all spins} & 
        \makecell{$\displaystyle  \frac{\big(\alpha X(k_1,k_2)\big)^{h_1+h_2+h_3}\,t^{\ta_1\ta_2\ta_3}_{h_1+h_2+h_3}}{\alpha\,(h_1+h_2+h_3-1)!}$  \vspace{.2cm} \\ $h_1+h_2+h_3>0$} &  \makecell{Moyal-SDYM \vspace{.2cm} \\   action($\Psi$) } \\
     \cmidrule(l r){1-3}
     \makecell{gl-chs($0$) \vspace{.2cm} \\    all spins} &  \makecell{$X(k_1,k_2)\,f^{\ta_1\ta_2\ta_3}$  \vspace{.2cm} \\ $h_1+h_2+h_3=1$} &  \makecell{SDYM \vspace{.2cm} \\  action($\Psi$) } \\
    \cmidrule(l r){1-3}
      \makecell{hsSDYM \vspace{.2cm} \\    $s\geq1$} & \makecell{$X(k_1,k_2)\,f^{\ta_1\ta_2\ta_3}$  \vspace{.2cm} \\ $s_1+s_2-s_3=1$} &   \makecell{SDYM \vspace{.2cm} \\  action($\Psi,\bar\Psi$) } \\ 
      
      \midrule\midrule \\

    \end{tabular}
  \end{threeparttable}
  \end{table}

%%%%%%%%%%%%%%%%%%%%%%%%%%%%%%%%%%%%%%%%%%
%%%%%%%%%%%%%%%%%%%%%%%%%%%%%%%%%%%%%%%%%%

\section{Double-copy structure and scattering amplitudes}
\label{sec:doublecopy}

In this section, building on ref.~\cite{Ponomarev:2017nrr}, we will see that the chiral higher-spin theories considered above inherit their double-copy structure from the Moyal-deformed theories. This leads to vanishing tree-level amplitudes beyond 3-point scattering.

Before proceeding, let us recall the double-copy structure of SDG and SDYM \cite{Monteiro:2011pc,Boels:2013bi}. These theories possess only a 3-point vertex, given by\footnote{The equations of motion for each theory were given in the first lines of \eqref{eq:SDGalpha} and \eqref{eq:SDYMalpha}.}
\[ 
\label{eq:3ptvert}
V_\text{SDG} = X(k_1,k_2)^2\,, \qquad V_\text{SDYM} = X(k_1,k_2) \,f^{\ta_1\ta_2\ta_3}\,,
\]
 where
\[
X(k_1,k_2):=k_{1w}k_{2u}-k_{1u}k_{2w}=-X(k_2,k_1) \,.
\]
These are bosonic vertices: notice that, due to momentum conservation, $X(k_i,k_j)$ is completely antisymmetric in $\{k_1,k_2,k_3\}$. The double-copy structure of these theories is based on the factorisation of the vertices into the structure constants of two Lie algebras: the self-dual-type `kinematic algebra' $X(k_1,k_2)$, and the colour Lie algebra in the case of $f^{\ta_1\ta_2\ta_3}$. The former is the algebra of area-preserving diffeomorphisms in the $(u,w)$ plane,
\[
\label{eq:LkMbracket}
[L_{k_1},L_{k_2}]= X(k_1,k_2)\,L_{k_1+k_2}\,,
\]
generated by Hamiltonian vector fields
\[
\label{eq:Lk}
L_{k}:=\{e^{i\,k\cdot x},\cdot\}=i \,e^{i\,k\cdot x}(k_u\partial_w-k_w\partial_u)= e^{i\,k\cdot x}\,k_u \,\epsilon^+\!\cdot\partial
\]
associated to positive helicity. In particular, the rightmost equality defines a standard positive-helicity polarisation vector.

For all the theories considered in this paper, we will find the double-copy structure
\[
\label{eq:3ptgeneral}
V = X(k_1,k_2) \,C_{I_1I_2}{}^{I_3}\,,
\]
where $C_{I_1I_2}{}^{I_3}$ are the structure constants of a Lie algebra specific to the theory. This fact was known for many of the theories studied \cite{Monteiro:2011pc,Ponomarev:2017nrr,Chacon:2020fmr}, to which we will add the cases of Moyal-SDYM and the chiral higher-spin theory gl-chs($\alpha$), related via \eqref{eq:MSDYMalpha}. Our approach emphasises how the double-copy structure of chiral higher-spin theories follows from the Moyal deformation.

\subsection{Kinematic algebras in Moyal-SDG}

The 3-point vertex in Moyal-SDG is 
\[
\label{eq:3ptvertMSDG}
V_\text{Moyal-SDG} = X(k_1,k_2)\, X^M(k_1,k_2)\,,
\]
where
\[
\label{eq:XM}
X^{M}(k_{1},k_{2}) := \frac1{\alpha} \sinh\big(\alpha\, X(k_{1},k_{2})\big) \,.
\]
So the double-copy structure is that we have one copy of $X(k_1,k_2)$ and one copy of its Moyal deformation $X^M(k_1,k_2)$ \cite{Chacon:2020fmr}. The former arises from a Poisson bracket and the latter arises from the associated Moyal bracket.\footnote{The counterpart of the generators \eqref{eq:Lk} in the deformed algebra is \,${ L_k^M}:=\{e^{i\,k\cdot x},\cdot\}^M= \frac{e^{i\,k\cdot x}}{\alpha} \sinh\big(\alpha\, k_u\,\epsilon^+\!\cdot \partial\big)$\,. Alternatively, the generators ${ E_k^M}:= \frac{e^{i\,k\cdot x}}{2\alpha} \exp\big(\alpha\, k_u\,\epsilon^+\!\cdot \partial\big)$ lead to the same structure constants. The latter choice is singular if we take $\alpha\to0$.}

Consider the Jacobi identity for the Moyal Lie bracket,
\[
\sinh\big(\alpha\, X(k_{1},k_{2})\big)\sinh\big(\alpha\, X(k_{1}+k_{2},k_{3})\big) + \text{cyc}(123) =0\,,
\]
where $\text{cyc}(123)$ denotes the other two cyclic permutations of legs 123. Introducing $k_4$ such that $\sum_{i=1}^4 k_i=0$, the expansion in $\alpha$ is
\begin{align}
0 & = \sinh\big(\alpha\, X(k_{1},k_{2})\big)\sinh\big(\alpha\, X(k_{3},k_{4})\big) + \text{cyc}(123) \nonumber \\
& = \sum_{\substack{\Lambda\geq 4 \\ \text{even\,} \Lambda}} \alpha^{\Lambda-2} \left(\,
\sum_{\substack{n=1 \\ \text{odd\,} n}}^{\Lambda-2}\,
\frac{X(k_1,k_2)^{\Lambda-2-n}}{(\Lambda-2-n)!}\; \frac{X(k_3,k_4)^{n}}{n!} \;+\; \text{cyc}(123)\right) .
\label{eq:jacMSDG}
\end{align}

Now, for the chiral higher-spin theory chs($\alpha$), with action \eqref{eq:actionchs}, the vertex is
\[
\label{eq:3ptvertchs}
V_{\text{chs}(\alpha)} = X(k_1,k_2) \cdot \frac1{\alpha}\, \frac{\big(\alpha X(k_1,k_2)\big)^{h_1+h_2+h_3-1}}{(h_1+h_2+h_3-1)!} = X(k_1,k_2) \,\tilde V_{\text{chs}(\alpha)} \,,
\]
where the helicities $h_i$ are taken to be incoming with respect to the vertex.
In this expression, we already show the two copies of structure constants, as originally identified in \cite{Ponomarev:2017nrr}. This should be compared to \eqref{eq:3ptvertMSDG}. The Jacobi relation for the algebra associated to $\tilde V_{\text{chs}(\alpha)}$ takes the form
\begin{align}
& \sum_{h_I}{}'\;  \frac{X(k_1,k_2)^{h_1+h_2+h_I-1}}{(h_1+h_2+h_I-1)!}\; \frac{X(k_3,k_4)^{h_3+h_4-h_I-1}}{(h_3+h_4-h_I-1)!} \;+\; \text{cyc}(123) \nonumber \\
&= \frac1{2\,(\Lambda-2)!}\, \Big[ \big(X(k_1,k_2)+X(k_3,k_4)\big)^{\Lambda-2} - \big(X(k_1,k_2)-X(k_3,k_4)\big)^{\Lambda-2}\Big] \,+\, \text{cyc}(123) \nonumber \\
&= 0\,,
\label{eq:jacchs}
\end{align}
where \,$\Lambda:=\sum_{i=1}^4 h_i$\, must be even, and moreover must satisfy $\Lambda\geq4$ because the sum over helicities in each vertex must be positive and, due to Bose symmetry, even. Momentum conservation leads to a cancellation in pairs in the final equality. The sum over $h_I$ in the first line is restricted (hence the prime) so that both exponents are positive and odd; so the `odd part' of the binomial formula is used in the first equality. The point is that the first line of \eqref{eq:jacchs} is identical to the coefficient with $\Lambda=\sum_{i=1}^4 h_i$ in the last line of \eqref{eq:jacMSDG}. The Jacobi relation for $X^{M}(k_{1},k_{2})$ encodes the Jacobi relation for $\tilde V_{\text{chs}(\alpha)}$, as noticed already in \cite{Ponomarev:2017nrr}. This follows from
\[
X^{M}(k_{1},k_{2})  =  \sum_{\sigma\geq1}\, \frac{\big(\alpha X(k_1,k_2)\big)^{2\sigma-1}}{\alpha(2\sigma-1)!} =
\sum_{\sigma\geq1}\, \tilde V_{\text{chs}(\alpha)}(h_1+h_2+h_3=2\sigma)\,. 
\]
The conclusion here is that the double-copy structure of Moyal-SDG encodes the double-copy structure of the associated chiral higher-spin theory.

For the simpler chiral higher-spin theory chs(0), where we have the restriction $h_1+h_2+h_3=2$, the double-copy structure of the vertex is the same as for SDG, in \eqref{eq:3ptvert}. Indeed, the expression for the vertex is the same, only the particle content is extended. The same conclusion applies to the theory hsSDG, with action \eqref{eq:actionhsSDG}, which is obtained from chs(0) by keeping only spins $s_i\geq2$ and vertices of the form $(+^{s_1}+^{s_2}-^{s_3})$.

\subsection{Kinematic and colour-kinematic algebras in Moyal-SDYM}

The 3-point vertex in Moyal-SDYM, with equation of motion \eqref{eq:MSDYMa}, is 
\[
\label{eq:3ptvertMSDSYM}
V_\text{Moyal-SDYM} = X(k_1,k_2) \Big( \! \cosh\big(\alpha X(k_1,k_2)\big)\,  f^{\ta_1\ta_2\ta_3}  +  \sinh\big(\alpha X(k_1,k_2)\big)\, d^{\ta_1\ta_2\ta_3}   \Big)\,.
\]
The double-copy structure is similar to the previous cases, with the Jacobi relation for the `right algebra' taking the form\footnote{Generators for the algebra are provided by ${ E_k^\ta{}^M}:= {e^{i\,k\cdot x}} T^\ta \exp\big(\alpha\, k_u\,\epsilon^+\!\cdot \partial\big)$. Notice that there is a restriction on the colour group, which we take to be U(N). }
\begin{align}
& f^{\ta_1\ta_2\tb}f^{\tb\ta_3\ta_4}\cosh\big(\alpha X(k_1,k_2)\big)\cosh\big(\alpha X(k_3,k_4)\big) \nonumber \\
+\,& f^{\ta_1\ta_2\tb}d^{\tb\ta_3\ta_4}\cosh\big(\alpha X(k_1,k_2)\big)\sinh\big(\alpha X(k_3,k_4)\big) \nonumber \\
+\,& d^{\ta_1\ta_2\tb}f^{\tb\ta_3\ta_4}\sinh\big(\alpha X(k_1,k_2)\big)\cosh\big(\alpha X(k_3,k_4)\big) \nonumber \\
+\,& d^{\ta_1\ta_2\tb}d^{\tb\ta_3\ta_4}\sinh\big(\alpha X(k_1,k_2)\big)\sinh\big(\alpha X(k_3,k_4)\big) \nonumber \\
+\,& \text{cyc}(123) =0\,.
\label{eq:jacMSDYM}
\end{align}
This is an interesting instance of the double copy. Instead of the usual factorisation between colour and kinematics as in \eqref{eq:3ptvert}, one of the algebras mixes colour and kinematics! In the limit $\alpha\to0$, we recover the usual colour Lie algebra. This type of deformation of the colour algebra appears in non-commutative Yang-Mills theory, e.g.~\cite{Armoni:2000xr}; see \cite{Huang:2010fc} for its appearance in a context close to ours.\footnote{
For a different recent appearance of $d^{\ta\tb\tc}$ in the double-copy literature, see \cite{Carrasco:2021ptp,Carrasco:2022jxn}.}

The associated gluonic chiral higher-spin theory gl-chs($\alpha$), with action \eqref{eq:actionglchs}, has the vertex
\[
\label{eq:3ptvertglchs}
V_{\text{gl-chs}(\alpha)} = X(k_1,k_2) \cdot \frac{t^{\ta_1\ta_2\ta_3}_{h_1+h_2+h_3}\,\big(\alpha X(k_1,k_2)\big)^{h_1+h_2+h_3-1}}{(h_1+h_2+h_3-1)!} = X(k_1,k_2) \,\tilde V_{\text{gl-chs}(\alpha)} \,,
\]
Similarly to the previous subsection, the relation \eqref{eq:jacMSDYM} can be checked to encode in its $\alpha$-expansion the Jacobi relation for $\tilde V_\text{gl-chs}$.

For the simpler gluonic chiral higher-spin theory gl-chs(0), where we have the restriction $h_1+h_2+h_3=1$, the double-copy structure of the vertex is the same as for SDYM, in \eqref{eq:3ptvert}. The same conclusion applies to the theory hsSDYM, with action \eqref{eq:actionhsSDYM}, which is obtained from gl-chs(0) by keeping only spins $s_i\geq1$ and vertices of the form $(+^{s_1}+^{s_2}-^{s_3})$.

\subsection{Comment on 3-point scattering amplitudes}
\label{subsec:3ptamp}

We now discuss the interpretation of 3-point scattering amplitudes, focusing on Moyal-SDG and $\text{chs}(\alpha)$. Consider the chiral higher-spin vertex
\[
\label{eq:3ptvertchs2}
V_{\text{chs}(\alpha)} = \frac{\big(\alpha X(k_1,k_2)\big)^{h_1+h_2+h_3}}{\alpha^2(h_1+h_2+h_3-1)!} \,.
\]
In our gauge choices and conventions, the 3-point amplitude is unhelpfully the same for any triplet of external states with the same $h_1+h_2+h_3$. We can backtrack slightly, and introduce the two following elements to compute amplitudes in the spinor helicity-formalism: we define\footnote{We follow the notation of refs.~\cite{Boels:2013bi,Monteiro:2022lwm}.}
\[
X(k_1,k_2)= \langle \eta | k_1k_2 | \eta\rangle\,,
\]
and introduce the helicities factor for $n$ external particles
\[
\prod_{i=1}^n\, (-\langle \eta i \rangle^{-2})^{h_i}\,.
\]
For massless particles, $X(k_1,k_2)=\langle \eta 1\rangle [12] \langle 2 \eta\rangle$. The rules here reduce to the previous ones for a certain choice of the spinors. We can exemplify the rules by checking that the 3-point amplitude is, up to a constant factor,
\begin{align}
-\frac{X(k_1,k_2)^{h_1+h_2+h_3}}{\prod_{i=1}^3 \langle \eta i \rangle^{2h_i}} & = 
-\frac{X(k_1,k_2)^{h_1+h_2-h_3}X(k_2,k_3)^{h_2+h_3-h_1}X(k_3,k_1)^{h_3+h_1-h_2}}{\prod_{i=1}^3 \langle \eta i \rangle^{2h_i}} \nonumber \\
& =  [12]^{h_1+h_2-h_3}[23]^{h_2+h_3-h_1}[31]^{h_3+h_1-h_2}\,,
\end{align}
as expected. Notice that, at 3 points, on-shell kinematics requires complexified momenta, so the amplitude has no support on real momenta in Lorentzian signature.

Now, the basic observation is that\footnote{See also related discussions in \cite{Ponomarev:2017nrr,Ponomarev:2022atv,Monteiro:2022lwm,Bu:2022iak}.}
\[
\label{eq:3ptvertMSDGsumchs}
V_\text{Moyal-SDG}  =  \sum_{\sigma\geq1}\, \frac{\big(\alpha X(k_1,k_2)\big)^{2\sigma}}{\alpha^2(2\sigma-1)!} =
\sum_{\sigma\geq1}\, V_\text{chs($\alpha$)}(h_1+h_2+h_3=2\sigma)\,. 
\]
This exhibits the breaking of Lorentz symmetry in Moyal-SDG that arises from the ($u,w$) plane of non-commutativity that is associated to $X(k_1,k_2)$. Notice that each term in the sum over $\sigma$ can give rise to a Lorentz-invariant 3-point amplitude if multiplied by appropriate external helicity factors (i.e.~such that $h_1+h_2+h_3=2\sigma$). There is no way, however, for the sum of these terms to exhibit Lorentz invariance --- understood as independence on $|\eta\rangle$ and having homogeneous little-group scaling in each particle. 

Still, the Lorentz non-invariance of Moyal-SDG is special, as the vertex generates vertices of the Lorentz-invariant theory chs($\alpha$). To see how it is special, we can consider the two conditions of Lorentz invariance in these theories with only 3-point light-cone-gauge vertices. The first condition is Lorentz invariance of the 3-point amplitudes, which is satisfied by chs($\alpha$) but not by Moyal-SDG for the reasons discussed above. The second is the `gluing' of 3-point vertices to form higher-point amplitudes (which turn out to vanish) in accordance with the closure of the Lorentz algebra, for which the numerical coefficient in each vertex is important; see e.g.~\cite{Ponomarev:2016lrm}. These numerical coefficients are constrained by the double-copy structure, in particular by the Jacobi relation \eqref{eq:jacchs} for the algebra $\tilde V_{\text{chs}(\alpha)}$. Moyal-SGD has a double-copy structure related to that of chs($\alpha$), as illustrated in \eqref{eq:jacMSDG}, so in this sense it satisfies the second condition for Lorentz invariance, while failing the first.

\subsection{Vanishing tree-level amplitudes}
\label{subsec:vanishing}

All the theories seen here have a 3-point vertex of the form \eqref{eq:3ptgeneral}. This subsection has a single piece of information: following the argument given in \cite{Monteiro:2022lwm}, based on the double copy, the fact that all these theories have the same `left algebra' associated to $X(k_1,k_2)$ means that $n$-point tree-level scattering amplitudes for $n>3$ vanish. For clarity, this is an on-shell statement. In the case of the gluonic higher-spin theory gl-chs($\alpha$), the vanishing was shown already in \cite{Skvortsov:2018jea}, based on Berends-Giele recursion for colour-ordered amplitudes. For the non-gluonic theory chs($\alpha$), the amplitude was shown to vanish at 4 points in \cite{Ponomarev:2016lrm}, and our observation provides an all-multiplicity proof. These proofs apply for generic kinematics. See \cite{Ponomarev:2022atv} for a recent discussion of amplitudes supported on special (complexified) kinematics with vanishing Mandelstam variables, which occurs also without higher spins \cite{Witten:2003nn}.

%%%%%%%%%%%%%%%%%%%%%%%%%%%%%%%%%%%%%%%%%%
%%%%%%%%%%%%%%%%%%%%%%%%%%%%%%%%%%%%%%%%%%

\section{Celestial chiral algebras}
\label{sec:celestial}

Celestial holography aims to interpret scattering amplitudes in four spacetime dimensions as correlation functions of a conformal-type field theory on the two-dimensional celestial sphere \cite{Cheung:2016iub,Pasterski:2016qvg,Pasterski:2017kqt,Pasterski:2017ylz}. The external states in a scattering amplitude correspond to operator insertions on the celestial sphere. As the figure below illustrates, the singular part of the operator product expansion (OPE) between two such operators is fixed by the collinear behaviour of the amplitudes \cite{Fan:2019emx,Pate:2019lpp}.
\begin{center}
\includegraphics[width=2.5cm]{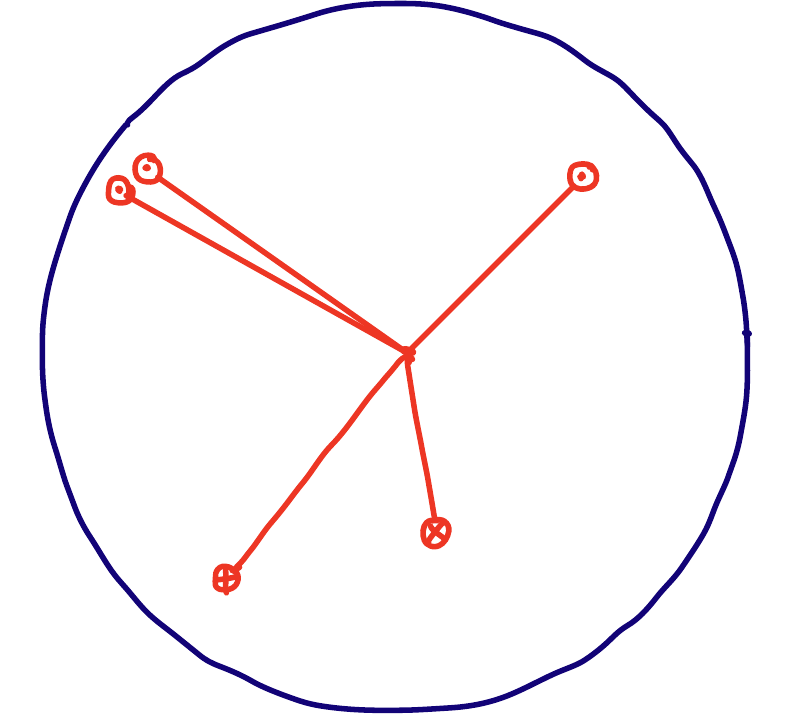}
\end{center}

In this section, we will use the momentum-space approach (and conventions) of \cite{Monteiro:2022lwm} to obtain the chiral celestial operator product expansions relevant to the theories considered in this paper. Let us review this approach and its results. We use the following parametrisation of massless momenta: 
\[
\label{eq:kin}
k_{A\dot A}=\lambda_A\tilde\lambda_{\dot A}\,,
\quad \lambda_A=(1,z) \,, \quad \tilde\lambda_{\dot A}=(k_u,k_w) \,,
\]
so that the holomorphic coordinate on the celestial sphere is
\[
z=\frac{k_{\bar w}}{k_u}=\frac{k_v}{k_w}\,.
\]
With this choice, we identify the anti-holomorphic spinor bracket with the self-dual kinematic algebra,
 \[
[12]:=\epsilon^{\dot A \dot B}\,\tilde\lambda_{1\,\dot A}\,\tilde\lambda_{2\,\dot B} = X(k_1,k_2) \,,
\]
and we can write the Mandelstam variables as
\[
s_{12} :=(k_1+k_2)^2= (z_1-z_2)\, { [12]} \,.
\]

Taking a `holomorphic collinear limit' $z_1\to z_2$, we obtain for the OPE in SDYM (which matches the $++$ OPE in full Yang-Mills)
\[
\label{eq:OPESDYM}
{\mathcal O}^{\ta_1}_+(k_1)\,\, {\mathcal O}^{\ta_2}_+(k_2) \,\sim\, \frac{{ X(k_1,k_2)} \,{ f^{\ta_1\ta_2\ta_3}}}{s_{12}} \,\,{\mathcal O}^{\ta_3}_+ (k_1+k_2)\,=\, \frac{{ f^{\ta_1\ta_2\ta_3}}}{z_1-z_2}\,\, {\mathcal O}^{\ta_3}_+(k_1+k_2)\,.
\]
The gravity counterpart is
\[
\label{eq:OPESDG}
{\mathcal O}_+(k_1)\,\, {\mathcal O}_+(k_2) \,\sim\, \frac{{ X(k_1,k_2)^2} }{s_{12}} \,\,{\mathcal O}_+ (k_1+k_2)\,=\, \frac{{ X(k_1,k_2)}}{z_1-z_2}\,\, {\mathcal O}_+(k_1+k_2)\,.
\]
The double-copy structure matches that of the celestial OPEs \cite{Monteiro:2022lwm,Guevara:2022qnm}: the $X$ algebra is associated to the chirality of the OPEs, while the second algebra (colour for SDYM and $X$ for SDG) provides the structure constants of the OPEs, ensuring associativity at tree level.

Let us remark that different points of view can be taken regarding the OPEs in SDYM and SDG. One is that we see SDYM and SDG as the self-dual sectors of full Yang-Mills and gravity, respectively, with their OPEs characterising the positive-helicity collinear behaviour of tree amplitudes in the full theories (since tree amplitudes beyond 3 points vanish in the self-dual sectors). Another point of view is to study the self-dual theories themselves, without consideration of full Yang-Mills and gravity. In this case, the OPEs are still meaningful because they are determined only by the 3-point amplitudes, which are non-vanishing in SDYM and SDG.\footnote{The interpretation in terms of collinear behaviour turns out to apply too within the self-dual theories, if we consider the one-loop amplitudes, which are very special. See the loop-level discussion in section~\ref{sec:loop}.}

The `soft-generator' OPEs are obtained from the OPEs above by taking the soft limit at fixed $z$, corresponding to $\tilde\lambda_{\dot A}=(k_u,k_w)\to(0,0)$. The operators are expanded as
\[
\label{eq:softexp}
{\mathcal O}^\ta_+(k)=\sum_{a,b=0}^\infty \frac{(ik_u)^a}{a!}\,\frac{(ik_w)^b}{b!}\;{ \varpi^\ta_{a,b}}(z)\,,
\quad \:\:
{\mathcal O}_+(k)=\sum_{a,b=0}^\infty \frac{(ik_u)^a}{a!}\,\frac{(ik_w)^b}{b!}\;{ \varpi_{a,b}}(z)\,.
\]
The soft generators lie in the `wedge' $a,b\geq0$.
They lead to the soft OPEs for SDYM,
\[
\label{eq:sOPESDYM}
\varpi^{\ta_1}_{a,b}(z_1)\, \varpi^{\ta_2}_{c,d}(z_2) \,\sim\, \frac{f^{\ta_1\ta_2\ta_3}}{z_1-z_2}\,\,\varpi^{\ta_3}_{a+c,b+d}(z_2)\,,
\]
and for SDG,
\[
\label{eq:sOPESDG}
\varpi_{a,b}(z_1)\, \varpi_{c,d}(z_2) \,\sim\, \frac{{ ad-bc}}{z_1-z_2}\,\,\varpi_{a+c-1,b+d-1}(z_2)\,.
\]
We may also expand the soft generators in Laurent modes,
\[
\label{eq:Laurent}
\varpi^{\ta}_{a,b}(z) = \sum_{n\in{\mathbb Z}} \, \frac{\varpi^{\ta}_{a,b\,;\,n}}{z^{n+1}}\,,
\qquad
\varpi_{a,b}(z) = \sum_{n\in{\mathbb Z}} \, \frac{\varpi_{a,b\,;\,n}}{z^{n+1}}\,.
\]
The SDYM modes satisfy the algebra
\[
\label{eq:msOPESDYM}
[\, \varpi^{\ta_1}_{a,b\,;\,n}\,,\, \varpi^{\ta_2}_{c,d\,;\,m} \,] =f^{\ta_1\ta_2\ta_3} \,\varpi^{\ta_3}_{a+c,b+d\,;\,n+m}\,,
\]
while for SDG we have
\[
\label{eq:msOPESDG}
[\, \varpi_{a,b\,;\,n}\,,\, \varpi_{c,d\,;\,m} \,] = (ad-bc) \,\varpi_{a+c-1,b+d-1\,;\,n+m}\,.
\]
For SDYM, we have an affine Kac-Moody algebra with level zero, which was identified in \cite{Fan:2019emx,Pate:2019lpp}; see also \cite{He:2015zea} for previous related work. For SDG, we have the loop algebra of the wedge subalgebra of $w_{1+\infty}$, which was identified in \cite{Strominger:2021lvk}\footnote{The translation between the conventions in \cite{Strominger:2021lvk} and ours is \,$w^p_m(z)=\frac1{2}\,\varpi_{p-1+m\,,\,p-1-m}(z)$\,.} building on \cite{Guevara:2021abz}; see the twistor derivation and historical discussion in \cite{Adamo:2021lrv}.\footnote{Despite the simplicity of these algebras of soft generators, the status of the latter as local operators in a celestial conformal field theory was questioned recently in \cite{Ball:2022bgg}.}

\subsection{Moyal-SDG and $LW_{\wedge}$}

We review now the case of Moyal-SDG, already analysed in \cite{Monteiro:2022lwm,Bu:2022iak}. From the vertex \eqref{eq:3ptvertMSDG}, we have
\[
\label{eq:OPEMSDG}
{\mathcal O}_+(k_1)\,\, {\mathcal O}_+(k_2) \,\sim\, \frac{{ X(k_1,k_2)}X^M(k_1,k_2) }{s_{12}} \,\,{\mathcal O}_+ (k_1+k_2)\,=\, \frac{{ X^M(k_1,k_2)}}{z_1-z_2}\,\, {\mathcal O}_+(k_1+k_2)\,.
\]
Expanding as in \eqref{eq:softexp}, we obtain
\begin{align}
\label{eq:sOPEMSDG}
& \varpi_{a,b}(z_1)\, \varpi_{c,d}(z_2) \,\sim \frac1{z_1-z_2}\; \times\nonumber \\
& \quad  \sum_{s\geq0}\frac{\alpha^{2s}}{(2s+1)!}\sum _{j=0}^{2s+1}(-1)^{j}\binom{2s+1}{j}
\,
[a]_{2s+1-j} [b]_j [c]_j [d]_{2s+1-j}
\;{ \varpi_{a+c-1-2s,b+d-1-2s}}(z_2)\,,
\end{align}
which is isomorphic to the loop algebra of the wedge subalgebra of $W_{1+\infty}$. $W_{1+\infty}$, introduced in \cite{Pope:1989ew,Pope:1989sr,Pope:1990kc} and reviewed in \cite{Pope:1991ig,Shen:1992dd}, is a higher-spin (in the 2D notion of spin) extension of the Virasoro algebra; the latter is generated by the Laurent modes of the (spin-2) chiral stress-energy tensor, whereas $W_{1+\infty}$ is an extension to generators of all spins$\geq1$. The isomorphism in the wedge subalgebra between \eqref{eq:sOPEMSDG} and $W_{1+\infty}$ follows from the results of \cite{Fairlie:1990wv}, and was well explained in \cite{Bu:2022iak}, where $LW_{\wedge}$ was used to denote the loop algebra in this class of isomorphic wedge algebras.  For $\alpha=0$, we recover the loop algebra of the wedge subalgebra of $w_{1+\infty}$ as in \eqref{eq:sOPESDG}. Refs.~\cite{Mago:2021wje,Ren:2022sws} discussed related deformations of the SDG chiral algebra.

\subsection{chs($\alpha$)}

We introduce now the celestial algebra of the chiral higher-spin theory chs($\alpha$), which is related to Moyal-SDG. We now have the vertex \eqref{eq:3ptvertchs}, which leads to the OPE
\[
\label{eq:OPEchs}
{\mathcal O}_{h_1}(k_1)\,\, {\mathcal O}_{h_2}(k_2) \,\sim\, 
\frac1{z_1-z_2} \sum_{\substack{h_3 \\ \text{even}\;\Sigma h>0}} \!  \frac{\big(\alpha\, X(k_1,k_2)\big)^{\Sigma h-1}}{\alpha(\Sigma h-1)!}\,\, {\mathcal O}_{-h_3}(k_1+k_2)\,,
\]
where \,$\Sigma h:=h_1+h_2+h_3$\,. The OPE is associative due to the double-copy structure of the vertex as before, with the Jacobi relation now corresponding to \eqref{eq:jacchs}. This proves an observation made in \cite{Ren:2022sws}, where it was noticed based on low-spin examples that the vertices of Metsaev that define chs($\alpha$) lead to associativity.

The soft expansion of the OPE leads to the wedge-type algebra
\begin{align}
& \varpi_{a,b}^{h_1}(z_1)\, \varpi_{c,d}^{h_2}(z_2) \,\sim \frac1{z_1-z_2}\sum_{\substack{h_3 \\ \text{even}\;\Sigma h>0}} \! \frac{\alpha^{\Sigma h-2}}{(\Sigma h-1)!} \;\times\nonumber \\
& \quad  \sum _{j=0}^{\Sigma h-1}(-1)^{j}\binom{\Sigma h-1}{j}
\,
[a]_{\Sigma h-1-j} [b]_j [c]_j [d]_{\Sigma h-1-j}
\;{ \varpi_{a+c+1-\Sigma h,b+d+1-\Sigma h}^{-h_3}}(z_2)\,.
\end{align}

\subsection{hsSDG}

For the theory hsSDG in \eqref{eq:actionhsSDG}, the vertex matches that of SDG but admits higher spins ($s\geq2$) such that $s_1+s_2-s_3=2$. This leads to an OPE that is a straightforward extension of the SDG case,
\[
{\mathcal O}_{s_1}(k_1)\,\, {\mathcal O}_{s_2}(k_2) \,\sim\, 
\frac{X(k_1,k_2)}{z_1-z_2} \; {\mathcal O}_{s_1+s_2-2}(k_1+k_2)\,.
\]
The soft expansion leads to the wedge-type algebra
\[
 \varpi_{a,b}^{s_1}(z_1)\, \varpi_{c,d}^{s_2}(z_2) \,\sim\, \frac{ad-bc}{z_1-z_2}
\;{ \varpi_{a+c-1,b+d-1}^{s_1+s_2-2}}(z_2)\,.
\]

\subsection{Moyal-SDYM and deformed Kac-Moody algebra}

From the vertex \eqref{eq:3ptvertMSDSYM} of Moyal-SDYM, we obtain the OPE
\begin{align}
\label{eq:OPEMSDYM}
& {\mathcal O}_{+}^{a_1}(k_1)\,\, {\mathcal O}_{+}^{a_1}(k_2)  \,\sim\, \nonumber \\
&  \frac1{z_1-z_2}\,\Big(\!  \cosh\big(\alpha X(k_1,k_2)\big)\,  f^{\ta_1\ta_2\ta_3}  +  \sinh\big(\alpha X(k_1,k_2)\big)\, d^{\ta_1\ta_2\ta_3} \Big) \, {\mathcal O}_{+}^{a_3}(k_1+k_2)\,.
\end{align}
The associated soft OPE is the wedge-type algebra
\begin{align}
& \varpi_{a,b}^{\ta_1}(z_1)\, \varpi_{c,d}^{\ta_2}(z_2) \,\sim \frac1{z_1-z_2}\; \sum_{s=0}^\infty\; \frac{\alpha^s}{s!} \; t^{\ta_1\ta_2\ta_3}_{s+1}\; \times\nonumber \\
& \qquad\qquad  \sum _{j=0}^{s}(-1)^{j}\binom{s}{j}
\,
[a]_{s-j} [b]_j [c]_j [d]_{s-j}
\;{ \varpi_{a+c-s,b+d-s}^{\ta_3}}(z_2)\,,
\end{align}
where $t^{\ta\tb\tc}_{s'}$ was defined in \eqref{eq:deft}. This is a deformation of a Kac-Moody algebra analogous to the case of $LW_{\wedge}$ for Moyal-SDG. For $\alpha=0$, we recover the Kac-Moody algebra of SDYM in \eqref{eq:sOPESDYM}. Refs.~\cite{Mago:2021wje,Ren:2022sws} discussed related deformations of the SDYM chiral algebra.

\subsection{gl-chs($\alpha$)}

For the U(N)-valued chiral higher-spin theory gl-chs($\alpha$), which is related to Moyal-SDYM, the vertex \eqref{eq:3ptvertglchs} leads to the OPE
\[
\label{eq:OPEglchs}
{\mathcal O}_{h_1}^{\ta_1}(k_1)\,\, {\mathcal O}_{h_2}^{\ta_2}(k_2) \,\sim\, 
\frac1{z_1-z_2} \,\sum_{\substack{h_3 \\ \Sigma h>0}}   \frac{t^{\ta_1\ta_2\ta_3}_{\Sigma h}\,\big(\alpha X(k_1,k_2)\big)^{\Sigma h-1}}{(\Sigma h-1)!}  \,\, {\mathcal O}_{-h_3}^{\ta_3}(k_1+k_2)\,,
\]
where \,$\Sigma h:=h_1+h_2+h_3$\,. As in all the previous cases, the OPE is associative due to the double-copy structure of the vertex. The soft expansion of the OPE leads to the wedge-type algebra
\begin{align}
& \varpi_{a,b}^{h_1,\ta_1}(z_1)\, \varpi_{c,d}^{h_2,\ta_2}(z_2) \,\sim \frac1{z_1-z_2}\,\sum_{\substack{h_3 \\ \Sigma h>0}} \frac{\alpha^{\Sigma h-1}}{(\Sigma h-1)!} \; t^{\ta_1\ta_2\ta_3}_{\Sigma h}\; \times\nonumber \\
& \quad  \sum _{j=0}^{\Sigma h-1}(-1)^{j}\binom{\Sigma h-1}{j}
\,
[a]_{\Sigma h-1-j} [b]_j [c]_j [d]_{\Sigma h-1-j}
\;{ \varpi_{a+c+1-\Sigma h,b+d+1-\Sigma h}^{-h_3,\ta_3}}(z_2)\,.
\end{align}

\subsection{hsSDYM}

Finally, for the theory hsSDYM in \eqref{eq:actionhsSDYM}, the vertex matches that of SDYM but admits higher spins ($s\geq1$) such that $s_1+s_2-s_3=1$. This leads to an OPE that is a straightforward extension of the SDYM case,
\[
{\mathcal O}_{s_1}^{\ta_1}(k_1)\,\, {\mathcal O}_{s_2}^{\ta_2}(k_2) \,\sim\, 
\frac{f^{\ta_1\ta_2\ta_3}}{z_1-z_2} \; {\mathcal O}_{s_1+s_2-2}^{\ta_3}(k_1+k_2)\,.
\]
The soft expansion leads to the wedge-type algebra
\[
\varpi_{a,b}^{s_1,\ta_1}(z_1)\, \varpi_{c,d}^{s_2,\ta_2}(z_2) \,\sim\, \frac{f^{\ta_1\ta_2\ta_3}}{z_1-z_2}
\;{ \varpi_{a+c,b+d}^{s_1+s_2-1,\ta_3}}(z_2)\,.
\]

%%%%%%%%%%%%%%%%%%%%%%%%%%%%%%%%%%%%%%%%%%
%%%%%%%%%%%%%%%%%%%%%%%%%%%%%%%%%%%%%%%%%%

\section{Loop level}
\label{sec:loop}

The loop-level study of chiral higher-spin theories was initiated in \cite{Skvortsov:2018jea,Skvortsov:2020wtf,Skvortsov:2020gpn}, which focused on one-loop planar amplitudes in the theory gl-chs($\alpha$) with action \eqref{eq:actionglchs}. In this section, we will discuss loop amplitudes in the various chiral higher-spin theories we considered. Many statements are explicit or implicit in \cite{Skvortsov:2018jea,Skvortsov:2020wtf,Skvortsov:2020gpn} and also in \cite{Adamo:2022lah}, so we are just aiming to give a useful overview. The collinear behaviour of loop amplitudes is relevant to the quantum fate of the tree-level celestial chiral algebras that we discussed above. Obviously, if the loop amplitudes vanish, then the chiral algebra is exact. It is sufficient, however, that the collinear behaviour of a theory is preserved at loop level, as exemplified in \cite{Ball:2021tmb} for SDG, where $w_{1+\infty}$ is a perturbatively exact symmetry. At one loop, this is implied by the vanishing of the tree amplitudes under some assumptions, because the one-loop correction to the collinear splitting function multiplies the tree amplitude; see e.g.~\cite{Bu:2022iak} for a discussion concerning Moyal-SDG. Recent results indicate that, for generic theories, the celestial OPEs acquire loop corrections that violate the associativity of the chiral algebra \cite{Bhardwaj:2022anh,Costello:2022upu,Bittleston:2022jeq}. For the chiral theories of the type studied here, no such violation is known.

Let us start with the theories hsSDG \eqref{eq:actionhsSDG} and hsSDYM \eqref{eq:actionhsSDYM}, which have the simplest structure of vertices. They possess only vertices of the type $(+^{s_1}+^{s_2}-^{s_3})$, for spins $s_i\geq2$ in the case of hsSDG, and $s_i\geq1$ in the case of hsSDYM. As such, like SDG and SDYM, they are one-loop exact theories, because it is not possible to write down higher-loop diagrams with only vertices of that type. It may be convenient for some readers to describe the argument in detail. The number of loops is
\[
L=I-v+1\,,
\]
where $I$ is the number of internal lines and $v$ is the number of vertices. The argument only cares about the signs of the helicities, so we can base it on $(++-)$ vertices.
\begin{center}
\includegraphics[width=2.5cm]{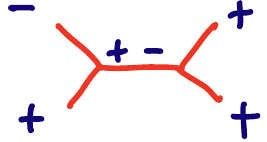}\qquad  \includegraphics[width=2.5cm]{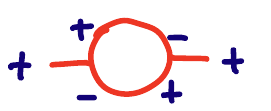}
\end{center}
Dressing the diagram with helicity signs as in the figures above, the plus and minus signs can be counted from either the lines (external and internal) or the $(++-)$ vertices:
\[
n_++I=2v\,,\qquad n_-+I=v\,.
\]
We obtain
\[
n_+=1-L+v\,,\qquad n_-=1-L\,.
\]
Hence, at tree level we have $n_-=1$, at one loop we have $n_-=0$, and there is no solution for $L>1$. Therefore, we need to consider only one loop. Let us consider first hsSDYM. Since at one loop the number of external particles is $n=n_+=v$, and since the sum over helicities in each vertex is $\Sigma h=1$, we have that $\sum_{i=1}^n h_i=n$. This requires that the external helicities, which must be positive, are $h_i=1$ in hsSDYM, just like in SDYM. For each one-loop diagram, any vertex that is not part of the loop is of the type $(+^{1}+^{1}-^{1})$ as in SDYM, while the vertices in the loop are of the type $(+^{1}+^{s}-^{s})$, and there is a sum over spins $s$ running in the loop. Therefore, the one-loop amplitudes in hsSDYM equal the ones in SDYM multiplied by a factor $\sum_{s\geq1}1$.\footnote{Notice that a factor of 2 accounting for $(+^{1}\pm^{s}\mp^{s})$ is already included in SDYM.} Following \cite{Skvortsov:2018jea}, we use the zeta-regularisation procedure of \cite{Beccaria:2015vaa} to identify that factor with $\zeta(0)=-\frac1{2}$. Hence, the non-vanishing one-loop amplitudes are \cite{Adamo:2022lah}
\[
{\mathcal A}^{(1)}_\text{hsSDYM}(1^1,2^1,\cdots,n^1)=-\frac1{2}\, {\mathcal A}^{(1)}_\text{SDYM}(1^+,2^+,\cdots,n^+)\,,
\]
where, on the left-hand side, the superscript indicates the helicity of each particle.
A similar story applies to hsSDG, with $h_i=2$ and a factor $\sum_{s\geq2}1$. We assume that the latter is regularised as $\zeta(0)-1=-\frac3{2}$. Hence, the non-vanishing one-loop amplitudes are
\[
{\mathcal A}^{(1)}_\text{hsSDG}(1^2,2^2,\cdots,n^2)=-\frac3{2}\, {\mathcal A}^{(1)}_\text{SDG}(1^+,2^+,\cdots,n^+)\,.
\]
A recent line of work \cite{Costello:2021bah,Costello:2022wso,Bittleston:2022nfr,Bittleston:2022jeq,Monteiro:2022nqt,Bu:2022dis} has described one-loop amplitudes in SDYM and SDG as resulting from an anomaly, as first suggested in \cite{Bardeen:1995gk}. The anomaly can be cancelled by coupling the gauge field to an `axion', leading to an anomaly free theory with vanishing loop amplitudes. The same can be done for hsSDYM and hsSDG, as recently pointed out in \cite{Adamo:2022lah}. Our discussion indicates that it is sufficient that the axion couples to spin 1 in hsSDYM and to spin 2 in hsSDG, since the higher spins do not contribute as external states to non-vanishing one-loop amplitudes. The coupling to the axion is then fixed by adjusting the coefficients that apply in SDYM and SDG to ones consistent with the two expressions for amplitudes above. Curiously, in the case of hsSDYM, if we truncate the theory to odd spins, and assume the zeta-function regularisation of $\sum_{\text{odd\,}s\geq1}1$ is $0$,\footnote{This corresponds to $\sum_{\text{odd\,}s\geq1}1=\sum_{n=1}^\infty (2n-1)^{-s}\to 0$ in the limit $s\to0$.} the one-loop amplitudes vanish. Following the discussions in \cite{Costello:2021bah,Costello:2022wso}, there should be no loop-level obstruction to directly uplifting this theory to twistor space. For hsSDG, if we truncate to even spins, the analogous zeta-function regularisation of $\sum_{\text{even\,}s\geq2}1$ leads instead to a factor  of $-\frac1{2}$.\footnote{This corresponds to $\sum_{\text{even\,}s\geq2}1=\sum_{n=1}^\infty (2n)^{-s}=2^{-s}\zeta(s)\to \zeta(0)=-\frac1{2}$ in the limit $s\to0$.}

Before proceeding to other theories, we note that hsSDG and hsSDYM have a collinear behaviour analogous to that of SDG and SDYM, and therefore their celestial algebras are quantum exact, similarly to the SDG case of $w_{1+\infty}$ \cite{Ball:2021tmb}. 

Let us move now to the chiral higher-spin theories chs(0) and gl-chs(0), obtained respectively from the actions \eqref{eq:actionchs} and \eqref{eq:actionglchs} after setting $\alpha=0$. Consider one loop. The vertices satisfy again $\Sigma h=2$ for chs(0) and $\Sigma h=1$ for gl-chs(0), so the external states must obey $\sum_{i=1}^n h_i=2n$ and $\sum_{i=1}^n h_i=n$ respectively. However, the external states do not individually have the same restriction as they did in hsSDG ($h_i=2$) and hsSDYM ($h_i=1$), because the vertices are not all of the type $(+^{s_1}+^{s_2}-^{s_3})$. Nevertheless, given an $m$-gon loop subdiagram (i.e.~$m=2$ is a bubble, $m=3$ is a triangle, etc), its `external' helicities obey the sum rules $\sum_{i=1}^m h_i=2m$ and $\sum_{i=1}^m h_i=m$ for chs(0) and gl-chs(0), respectively; and moreover, fixing a single helicity along the loop fixes them all, with the sum rule being the consistency condition. The conclusion is that the possibility of fixing a single helicity along the loop to any value still leads, for the amplitudes, to a factor representing the sum over helicities running in the loop. This factor is
\[
\label{eq:sumh0}
\sum_h 1=1+2\sum_{s\geq1}1=1+2 \zeta(0)=0\,,
\]
where the isolated 1 is the scalar contribution, and $2\sum_{s\geq1}1$ is the contribution from the other helicities. This vanishing factor multiplies in chs(0) and gl-chs(0) the one-amplitudes of hsSDG and hsSDYM, and there is also a helicity-dependent overall factor in each case. The point is that the vanishing factor multiplies a finite contribution, so the one-loop amplitudes vanish, which is the argument given in \cite{Skvortsov:2018jea,Skvortsov:2020wtf,Skvortsov:2020gpn}. Remarkably, if we truncate gl-chs(0) to admit only odd spins, we get zero again from $2\sum_{\text{odd\,}s\geq1}1=0$; and if we truncate chs(0) to admit only even spins, we get zero again from $1+2\sum_{\text{even\,}s\geq2}1=1+2(-\frac1{2})=0$. We conclude that the theories chs(0) and gl-chs(0) are actually simpler at one loop than hsSDG and hsSDYM.

At higher loops, however, the amplitudes in hsSDG and hsSDYM manifestly vanish due to the class of vertices, as we mentioned --- an argument that does not apply to chs(0) and gl-chs(0). To illustrate this point, we can start by recalling the story for SDYM. Its equation of motion admits two types of cubic action: one with a single field \cite{Leznov:1986mx,PARK1990287},
\[
\label{eq:SDYMbadaction}
S_{\Psi^3\,\text{`SDYM'}}(\Psi) = \int d^4 x \;\; \text{tr}\left(\frac1{2}\,\Psi\square \Psi + \frac{1}{3}\,\Psi[\partial_u \Psi,\partial_w \Psi]\right) \,,
\]
and one with two fields representing opposite helicities that was introduced later in \cite{Chalmers:1996rq},
\[
\label{eq:SDYMaction}
S_\text{SDYM}(\Psi,\bar\Psi) = \int d^4 x \;\; \text{tr}\; \bar\Psi \left(\square \Psi + [\partial_u \Psi,\partial_w \Psi]\right) \,.
\]
This second action has a ($++-$) vertex and is therefore one-loop exact. This is the correct action for SDYM as a quantum field theory, and it can be seen as a sector of the full Yang-Mills theory. The theories hsSDG and hsSDYM are analogous to this case, as it is clear in the way they are generated from SDG and SDYM in \eqref{eq:actionhsSDG} and \eqref{eq:actionhsSDYM}. The action \eqref{eq:SDYMbadaction}, on the other hand, admits diagrams with any number of loops. This is not a Lorentz-invariant theory: the propagator does not connect fields of opposite helicity, unless the field is a scalar, but then the vertex would unavoidably break Lorentz invariance because there is no gauge to be fixed. Surprisingly, this non-Lorentz-invariant theory does generate the Lorentz-invariant theory gl-chs(0) as seen in \eqref{eq:actionglchs} if we set $\alpha=0$. Given that the one-loop amplitudes vanish so beautifully due to a sum over helicities, it is tempting to conclude that the same will happen at higher loops. In fact, the sum \eqref{eq:sumh0} will still arise for an $m$-gon loop subdiagram with given `external' helicities joining the loop. However, it is not clear now whether such vanishing factors multiply a finite kinematic function; it was clear at one loop because of the relation to one-loop amplitudes in SDG and SDYM. If the kinematic functions diverge, perhaps the conclusion is still that the loop amplitudes vanish, but there is a worrying competition of regularisations (states running in the loop versus, say, dimensional regularisation). We will not pursue this further here.  

The `oldest' and most complicated chiral higher-spin theories are chs($\alpha$) and gl-chs($\alpha$) for $\alpha\neq0$, with actions \eqref{eq:actionchs} and \eqref{eq:actionglchs}. The gl-chs($\alpha$) case was studied in \cite{Skvortsov:2018jea,Skvortsov:2020wtf,Skvortsov:2020gpn}, where it was found that the planar one-loop amplitudes are again related to the planar SDYM amplitudes, with a helicities-dependent dressing factor; they still vanish in view of an overall factor giving the sum over states running in the loop, which vanishes in zeta regularisation, similarly to gl-chs(0). In fact, the arguments above still apply, because an $m$-gon loop (sub)diagram, with given `external' helicities and with given $\Sigma h$ for each vertex, will provide the sum \eqref{eq:sumh0} over states in the loop. Now, however, it is even harder to know whether such vanishing factors multiply a finite kinematic function, due to the higher tensorial rank of the numerators in a loop integrand. At one loop, this was checked explicitly in \cite{Skvortsov:2018jea,Skvortsov:2020wtf,Skvortsov:2020gpn} for planar gl-chs($\alpha$). For the non-planar part and for chs($\alpha$), we still have the unitarity-based argument that the one-loop amplitudes cannot have branch cuts because the tree amplitudes vanish, but this does not show that the amplitudes are finite. Beyond one loop, the status of chs($\alpha$) and gl-chs($\alpha$) is at least as unclear as the status of their $\alpha=0$ counterparts mentioned earlier: sums over states in the loops suggest that the amplitudes vanish, but the possibility of divergences in the loop integration should give us pause.

Finally, one may wonder about the relation to amplitudes in Moyal-SDG and Moyal-SDYM, seen respectively as theories of spin 2 and spin 1 particles with Lorentz-breaking higher-derivative corrections. Notice that the Moyal deformed self-dual equations of motion admit actions analogous to \eqref{eq:SDYMbadaction} or to \eqref{eq:SDYMaction}. In the latter case, the actions define one-loop exact theories. In either case, the one-loop amplitudes cannot have branch cuts because the tree amplitudes vanish. As for the relation to the chiral higher-spin theories, it arises from actions of the type \eqref{eq:SDYMbadaction}, as we saw in \eqref{eq:actionchs} and \eqref{eq:actionglchs}. Hence, {\it a priori}, the chiral higher-spin theories could have amplitudes of any loop order. We can consider the correspondence \eqref{eq:3ptvertMSDGsumchs} between Moyal-SDG and chs($\alpha$). It implies that a loop integrand in chs($\alpha$), with a specific choice of external helicities, is related to a particular power of $\alpha$ in the analogous Moyal-SDG integrand, up to the sum over states running in the loop and to an overall factor fixing the external helicity weights. However, the perturbative behaviour of (non-planar) theories on non-commutative spaces can be non-analytical in the deformation parameter \cite{Filk:1996dm,Minwalla:1999px}, so this may be a subtle question, and we will not pursue it here. For planar theories on non-commutative spaces, e.g.~planar Moyal-SDYM, the amplitudes at any loop order turn out to match those of the limiting commutative theories times a `phase', due to the form of the Moyal product at each vertex \cite{Filk:1996dm,Minwalla:1999px}. (An actual phase arises from defining our non-commutativity parameter $\alpha$ to be imaginary, which is also a common choice.) Hence, presumably the planar one-loop results of \cite{Skvortsov:2018jea,Skvortsov:2020wtf} for gl-chs($\alpha$), which is related to Moyal-SDYM, can be interpreted in the following manner: start with a planar one-loop SDYM amplitude; deform it into a Moyal-SDYM amplitude using the phase, whose argument is proportional to $\alpha$; expand that amplitude in $\alpha$; specify the desired external higher-spin states in the gl-chs($\alpha$) amplitude; pick up the corresponding term in the $\alpha$ expansion, which is selected by the value of the sum of helicities of the external states; multiply by the appropriate external helicities factor, as discussed in section~\ref{subsec:3ptamp}; and, finally, multiply also by the vanishing sum over states running in the loop \eqref{eq:sumh0}. Beyond one-loop (but still in the planar sector), we may expect to proceed in the same way by starting with `SDYM' and its Moyal deformation, whose actions appear in \eqref{eq:SDYMbadaction} and \eqref{eq:actionglchs}; notice that while they match SDYM and Moyal-SDYM at one loop, they may have non-vanishing higher-loop amplitudes. However, we come across the issue discussed above with the competition of regularisations: we would need to understand the integration over loop momenta in `SDYM'.
We will not pursue these questions further here.

%%%%%%%%%%%%%%%%%%%%%%%%%%%%%%%%%%%%%%%%%%
%%%%%%%%%%%%%%%%%%%%%%%%%%%%%%%%%%%%%%%%%%

\section{Conclusion}
\label{sec:conclusion}

We discussed various connections between chiral-type theories, namely self-dual gravity or Yang-Mills, their Moyal deformations, and chiral higher-spin theories. Table~\ref{tablechs} summarised some of the main results. We described a notion of theories `generating' other theories, which is realised in a simple manner in light-cone gauge. Indeed, there is a wonderful ambiguity in the use of light-cone gauge: if one is given a gauge-fixed equation, it may prove difficult to understand basic facts about its constituents, such as the spin of the fields; and yet, not only does the simplicity of the expressions --- particularly for chiral theories --- allow one to obtain crucial insights about the structure of a theory, but also the `ambiguity' can be fruitful in suggesting connections between theories. The very existence of chiral higher-spin theories was noticed first using light-cone-gauge methods  \cite{Metsaev:1991mt,Metsaev:1991nb,Ponomarev:2016lrm}. Another example is the straightforward off-shell realisation of the colour-kinematics duality in self-dual Yang-Mills theory \cite{Monteiro:2011pc}. We saw how all the chiral-type theories were akin to self-dual Yang-Mills, in that the double-copy structure of the vertex has one copy of the self-dual kinematic algebra and one copy of another (colour, kinematic or colour-kinematic) algebra. We also mentioned how this is sufficient to guarantee the vanishing of the tree-level amplitudes in all these theories, applying an argument from \cite{Monteiro:2022lwm} to new examples. (One may ask, in the context of the double copy, what happens if one uses a vertex that is the product of, say, two Moyal-deformed self-dual kinematic algebras. The amplitudes do not vanish, but the theory breaks Lorentz invariance, apparently without an interesting interpretation.)

Along with making these connections between theories, one of our main goals was to describe associativity-preserving deformations of OPE algebras arising in celestial holography, in particular the chiral algebras of soft modes of gluons and gravitons; see \cite{Himwich:2021dau,Mago:2021wje,Ren:2022sws} for related work. The case of Moyal-deformed self-dual gravity, which we reviewed, was discussed recently in \cite{Monteiro:2022lwm,Bu:2022iak} and is related to $W_{1+\infty}$. In this paper, we added the case of Moyal-deformed self-dual Yang-Mills, which gives a related deformation of a U(N) Kac-Moody algebra. We also presented here the (2D-)chiral algebras for the (4D-)chiral higher-spin theories. 

Finally, we gave an overview of loop amplitudes in these various theories. Light-cone gauge played a crucial role here again, allowing us to focus our arguments on physical degrees of freedom, thereby avoiding the need for Faddeev-Popov ghosts.

There are various open questions. Regarding chiral higher-spin theories, it would be interesting to consider extensions, such as the theory recently discussed in \cite{Adamo:2022lah,Tran:2022amg} and also supersymmetric extensions.

Regarding the celestial algebras, this paper builds on the results of \cite{Monteiro:2022lwm,Guevara:2022qnm} connecting the colour-kinematics duality to the associative structure of the tree-level celestial algebras. Indeed, it would be surprising if these were two fully distinct algebraic structures. It is important, however, to extend this connection to the full theories, i.e.~beyond the chiral sector. The simplest next step would be the MHV (next-to-self-dual) sector, where relevant OPEs have been recently worked out to all orders in the collinear expansion \cite{Adamo:2022wjo}. At loop level, the connection is harder to envisage, but its exploration may lead to lessons on both sides. We could also study in greater detail the various chiral-type theories discussed here even at tree level. We came across a deformation of a U(N) Kac-Moody algebra, and it would be interesting to study it further and to extend it to central terms and other parameters, similarly to the case of $W_{1+\infty}$.

As a final open question, the precise relations of some or all of the chiral-type theories considered here to the ${\mathcal N}=2$ strings remain to be fully uncovered; see e.g.~\cite{Ooguri:1991fp,Chalmers:2000bh,Lechtenfeld:2000nm}.

%%%%%%%%%%%%%%%%%%%%%%%%%%%%%%%%%%%%%%%%%%

\section*{Acknowledgements}

We are grateful to Alfredo Guevara, Henrik Johansson, Radu Roiban, Evgeny Skvortsov, Andrew Strominger and Tung Tran for discussions and suggestions. This work was supported by a Royal Society University Research Fellowship.

\bibliography{refs}
\bibliographystyle{JHEP}

\end{document}